\documentclass[twocolumn,prl]{revtex4}
\usepackage{graphicx}    
\usepackage{dcolumn}     
\usepackage{slashbox}     
\usepackage{bm}          
\newcommand{\Vec}[1]{\mbox{\boldmath$#1$}}
\begin{document}
\title{
Unconventional pairing originating from 
disconnected Fermi surfaces in superconducting LaFeAsO$_{1-x}$F$_x$}
\author{
Kazuhiko Kuroki$^1$, Seiichiro Onari$^2$, Ryotaro Arita$^{3}$\cite{arita},\\  
Hidetomo Usui$^1$, Yukio Tanaka$^2$, Hiroshi Kontani$^4$, and 
Hideo Aoki$^5$ 
}
\affiliation{
$^1$ Department of Applied Physics and Chemistry, 
The University of Electro -Communications, Chofu, Tokyo 182-8585, Japan}
\affiliation{$^2$ Department of Applied Physics, Nagoya University, 
Nagoya 464-8603, Japan and CREST-JST}
\affiliation{$^3$ RIKEN, 2-1 Hirosawa, Wako, Saitama 351-0198, Japan}
\affiliation{$^4$ Department of Physics,  
Nagoya University, Nagoya 464-8602, Japan} 
\affiliation{$^5$ Department of Physics,  
University of Tokyo, Tokyo 113-0033, Japan}
\date{\today}
\begin{abstract}
For a newly discovered iron-based high $T_c$ superconductor 
LaFeAsO$_{1-x}$F$_x$, we have constructed a minimal model, where 
inclusion of all the five Fe $d$ bands is found to be necessary.  
Random-phase approximation is applied to the model 
to investigate the origin of superconductivity. We conclude that 
the multiple spin fluctuation modes arising from 
the nesting across the disconnected Fermi 
surfaces realize an extended $s$-wave pairing, while 
$d$-wave pairing can also be another candidate.
\end{abstract}
\pacs{PACS numbers: }
\maketitle
Understanding the mechanism of unconventional superconductivity (SC) 
has been one of the most challenging problems in 
condensed-matter physics.  There is a renewed fascination with 
a recent discovery of SC in an 
iron-based superconductor LaFeAsO$_{1-x}$F$_x$,\cite{Hosono2} 
which is likely to provide a fresh avenue for such a challenge.
LaFeAsO belongs to the family of 
quaternary oxypnictides LnMPnO (Ln=La, Pr; M=Mn, Fe, Co, and Ni; 
Pn=P, As), which was originally fabricated by Zimmer {\it et al.} and 
Quebe {\it et al.}\cite{Zimmer,Quebe}
For this family of compounds, Kamihara {\it et al.} first reported 
that LaFePO exhibits SC with $T_c\simeq 3$K, which was raised to 
$T_c\simeq 7$K by F doping.\cite{Hosono1} 
SC has also been found in nickel-based compounds with the same 
structure.\cite{Hosono3}
Very recently, Kamihara {\it et al.} have come up with 
the discovery of SC in LaFeAsO$_{1-x}$F$_x$, where 
the F doping with $x \simeq 0.11$ leads to a remarkable $T_c\sim 26$K. 

The high value of $T_c$ itself, confirmed also 
by Chen {\it et al.},\cite{Chen} 
suggests a possibility of unconventional SC, 
but direct evidences are accumulating: 
A specific heat measurement in magnetic fields  
shows that the coefficient $\gamma$ 
displays a $\sqrt{H}$ behavior.\cite{Mu}
A point-contact conductance measurement exhibits  
spectra with a distinct zero-bias peak,\cite{Shan} 
suggestive of the presence of sign change in the gap function.
\cite{Hu,Tanaka1,Tanaka3,Lofwander} 
The starting material, LaFeAsO, is a bad metal 
with some anomaly in the resistivity 
around 100K.\cite{Hosono2} As the system becomes metallic 
upon F doping, the uniform susceptibility exhibits a Curie-Weiss 
behavior. Anomalies in the normal-state 
transport properties have also been reported for doped systems.\cite{Zhu}

Theoretically, first-principles band structure has been obtained for  
LaFePO\cite{Lebegue}, 
and more recently for LaFeAsO and related materials
\cite{Singh,Kotliar,Xu,Boeri}. These band structures are metallic with five 
pieces (sheets) of the Fermi surface in the undoped system, which 
contradicts with the experiment for the undoped LaFeAsO.\cite{Hosono2} 
However, a dynamical mean-field study 
shows that the electron correlation enhances the crystal field 
splitting, which leads to a 
band-semiconducting behavior in accord with the experiment.\cite{Kotliar} 
Local spin-density calculations for LaFeAsO show that the system 
is around the border between magnetic and nonmagnetic states, with 
a tendency toward ferromagnetism and 
antiferromagnetism.\cite{Singh,Xu}  It is also pointed out that the 
electron-phonon coupling 
is too weak to account for $T_c=26$K.\cite{Mu,Boeri} 

Given this background, the purpose of the present Letter 
is to first construct a microscopic 
electronic model for LaFeAsO$_{1-x}$F$_x$, which then serves as the basis 
for identifying the possible mechanisms why this material 
favors high-$T_c$.  
The minimal model has turned out to contain all the five Fe $d$ orbitals, 
to which we have applied the random-phase approximation (RPA) 
to solve the Eliashberg equation. 
We shall conclude that a peculiar Fermi surface consisting of 
multiple pockets and ensuing multiple 
spin-fluctuation modes 
realize an unconventional $s$-wave pairing, 
while $d$-wave pairing can also be another candidate.

\begin{figure}[b]
\begin{center}
\includegraphics[width=8.5cm,clip]{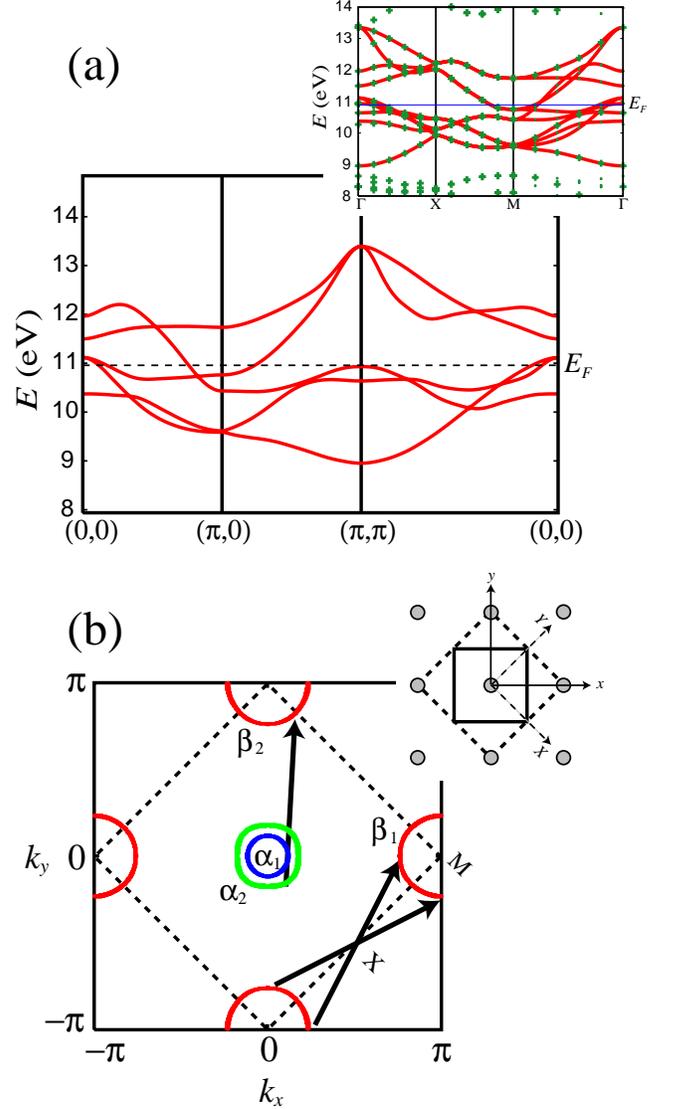}
\caption{(color online)
(a) The band structure of the five-band model 
in the unfolded BZ, where the 
inter-layer hoppings are included.  To compare with 
the ten-band model (red lines in the inset; the symbols are the 
present LDA results), note the original (dashed lines) and 
the unfolded (solid) BZ shown in Fig.(b).
(b) Fermi surface for $n=6.1$ (with the 
inter-layer hoppings ignored), with the arrows indicating the 
nesting vectors. 
Inset depicts the original (dashed) and 
reduced (solid) unit cell in real space. 
\label{fig1}}
\end{center}
\end{figure}

LaFeAsO has a tetragonal layered structure, in which Fe atoms are arrayed on a 
square lattice. 
Due to the tetrahedral coordination of As, 
there are two Fe atoms per unit cell.  
Each Fe layer is then sandwiched between LaO layers.
The experimentally determined lattice constants 
are $a=4.03552$\AA and $c=8.7393$\AA, with 
two internal coordinates $z_{La}=0.1415$ and $z_{As}=0.6512$.
We have obtained the band structure (Fig.\ref{fig1}(a) inset) with the 
Quantum-ESPRESSO package\cite{pwscf}, and then 
construct the maximally localized 
Wannier functions (MLWFs)\cite{MaxLoc}. 
These MLWFs, centered at the two Fe sites in the unit cell, 
have five orbital symmetries (orbital 1:$d_{3Z^2-R^2}$, 
2:$d_{XZ}$, 3:$d_{YZ}$, 4:$d_{X^2-Y^2}$,
5:$d_{XY}$, where $X, Y, Z$ refer to those for the original unit
cell).  
We can note that the two Wannier orbitals in 
each unit cell are equivalent in that each Fe atom has the same 
local arrangement of other atoms.
We can thus take a unit cell that 
contains only one orbital per symmetry by 
unfolding the Brillouin zone (BZ),\cite{comment}  
and we end up with an effective five-band model on a 
square lattice, where 
$x$ and $y$ axes are rotated by 
45 degrees from $X$-$Y$ (Fig.\ref{fig1}(b) inset), 
to which we refer for all the wave vectors hereafter. 
The in-plain 
hopping integrals $t(\Delta x, \Delta y, \Delta z=0;\mu,\nu)$ 
are displayed in table \ref{tab1}, 
where $[\Delta x, \Delta y]$ is the hopping vector, 
and $\mu$, $\nu$ label the five Wannier orbitals.
The on-site energies for the five orbitals are 
$(\varepsilon_1,\varepsilon_2,\varepsilon_3,\varepsilon_4,
\varepsilon_5)=(10.75,10.96,10.96,11.12,10.62)$ eV. 
With these effective hoppings and on-site energies, 
the in-plane tight-binding Hamiltonian is given in the form 
\begin{eqnarray}
&&H_0=\sum_{ij}\sum_{\mu\nu}\sum_\sigma
\left[t(x_i-x_j, y_i-y_j;\mu,\nu)c_{i\mu\sigma}^\dagger c_{j\nu\sigma}
\right.
\nonumber\\
&+&
\left.
t(x_j-x_i, y_j-y_i;\nu,\mu)c_{j\nu\sigma}^\dagger c_{i\mu\sigma}\right]
+\sum_{i\mu\sigma}\varepsilon_\mu n_{i\mu\sigma},
\end{eqnarray}
where $c_{i\mu\sigma}^\dagger$ creates an electron with spin $\sigma$ 
on the $\mu$-th orbital at site $i$, and 
$n_{i\mu\sigma}=c^\dagger_{i\mu\sigma}c_{i\mu\sigma}$.
We define the band filling $n$ as the number of electrons/number of sites
(e.g., $n=10$ for full filling). 
The doping level $x$ 
in LaFeAsO$_{1-x}$F$_x$ is related to the band filling as $n=6+x$.

\begin{table}
\begin{tabular}{c|rrrrrccc}
\backslashbox{\small($\mu$,$\nu$)}{\small $[\Delta x, \Delta y]$}
 &[1,0]&[1,1]&[2,0]&[2,1]&[2,2]&$\sigma_y$& $I$& $\sigma_d$\\ 
\hline
  (1,1) &  $-$0.7&      &$-$0.4&   0.2 &$-$0.1 &   +    & +   &+  \\
  (1,2) &  $-$0.8&      &      &       &       &$-$(1,3)& $-$ &$-$\\
  (1,3) &     0.8&$-$1.5&      &       &$-$0.3 &$-$(1,2)& $-$ &+  \\
  (1,4) &        &   1.7&      &       &$-$0.1 &   $-$  & +   &+  \\
  (1,5) &  $-$3.0&      &      &$-$0.2 &       &    +   & +   &$-$\\
  (2,2) &  $-$2.1&   1.5&      &       &       &+(3,3)  & +   &+  \\
  (2,3) &     1.3&      &   0.2&$-$0.2 &       &    +   & +   &$-$\\
  (2,4) &     1.7&      &      &   0.2 &       &+(3,4)  & $-$ &$-$\\
  (2,5) &  $-$2.5&   1.4&      &       &       &$-$(3,5)& $-$ &+  \\
  (3,3) &  $-$2.1&   3.3&      &$-$0.3 &   0.7 &+(2,2)  & +   &+  \\
  (3,4) &     1.7&   0.2&      &   0.2 &       &+(2,4)  & $-$ &+  \\
  (3,5) &     2.5&      &      &   0.3 &       &$-$(2,5)& $-$ &$-$\\
  (4,4) &     1.6&   1.2&$-$0.3&$-$0.3 &$-$0.3 &   +    & +   &+  \\
  (4,5) &        &      &      &$-$0.1 &       &  $-$   & +   &$-$\\
  (5,5) &     3.1&$-$0.7&$-$0.2&       &       &   +    & +   &+  \\
\end{tabular}
\caption{Hopping integrals $t(\Delta x, \Delta y; \mu, \nu)$ 
in units of 0.1eV. $[\Delta x, \Delta y]$ denotes the in-plain 
hopping vector, 
and $(\mu,\nu)$ the orbitals. $\sigma_y$, $I$, and 
$\sigma_d$ corresponds to $t(\Delta x, -\Delta y;\mu,\nu)$, 
$t(-\Delta x, -\Delta y;\mu,\nu)$, $t(\Delta y, \Delta x;\mu,\nu)$, 
respectively, where `$\pm$' and `$\pm (\mu',\nu')$' in the row of 
$(\mu,\nu)$ mean that the corresponding  
hopping is equal to $\pm t(\Delta x, \Delta y; \mu, \nu)$ and  
$\pm t(\Delta x, \Delta y; \mu', \nu')$, respectively. 
This table, combined with the relation 
$t(\Delta x, \Delta y; \mu, \nu) = t(-\Delta x, -\Delta y; \nu, \mu)$, 
gives all the in-plain hoppings $\geq 0.01$eV up to fifth 
neighbors.\label{tab1}}
\end{table}

In the obtained band structure in Fig.\ref{fig1}(a), 
we notice that the five bands are heavily 
entangled, reflecting strong hybridization (see table \ref{tab1}) 
of the five $3d$ orbitals, which is physically due to the tetrahedral 
coordination of As atoms around Fe.
Hence we conclude that the minimal electronic model 
requires all the five bands. In Fig.\ref{fig1}(b), 
the Fermi surface for $n=6.1$ (corresponding to $x=0.1$) 
obtained by ignoring the inter-layer hoppings 
is shown in the two-dimensional unfolded BZ.
The Fermi surface consists of 
four pieces (pockets in 2D):   
two concentric hole pockets (denoted as $\alpha_1$, $\alpha_2$) 
centered around $(k_x, k_y)=(0,0)$, two electron pockets 
around $(\pi,0)$ $(\beta_1)$ or $(0,\pi)$ $(\beta_2)$, respectively. 
$\alpha_i$ ($\beta_i$) corresponds to the 
Fermi surface around the $\Gamma$Z (MA) line (in the original BZ) 
in the first-principles band calculation.
\cite{Lebegue,Singh,Xu}

Having constructed the model, we now move on to the RPA calculation.
We again adopt the 2D model in which the 
inter-layer hoppings are neglected.\cite{comment4}
For the many-body part of the Hamiltonian,  
we consider the standard interaction terms that comprise 
the intra-orbital Coulomb $U$, the inter-orbital 
Coulomb $U'$, the Hund's coupling $J$, and the pair-hopping $J'$. 
All the calculation is done in the orbital representation.
Details of the multiorbital RPA calculation can be found in e.g.
ref.\cite{Yada,Takimoto}. The modification of the band structure due to 
the self-energy correction is not taken into account, 
on which we comment later.
In the present case, 
the Green's function is a $5\times 5$ matrix, while 
the spin and the orbital susceptibilities become $25\times 25$ matrices.
The Green's function and 
the effective pairing interactions, obtained from the susceptibilities,  
are plugged into the linearized Eliashberg equation, and 
the gap function in a $5\times 5$ matrix form
along with the eigenvalue $\lambda$ is obtained. 
$T_c$ corresponds to the temperature where $\lambda$ reaches unity.
$32\times32$  $k$-point meshes and 1024 Matsubara frequencies are taken. 
We find that 
the spin fluctuations dominate over orbital fluctuations as far as 
$U>U'$, so we focus on the spin susceptibility.
We denote the largest eigenvalue of the spin 
susceptibility matrix for $i \omega_n=0$ 
as $\chi_s(\Vec{k})$. 
The gap function matrix at the lowest Matsubara frequency 
is transformed into the band representation by 
a unitary transformation, and its 
diagonal element for band $i$ is denoted as 
$\phi_i(\Vec{k})$.
\begin{figure}[b]
\begin{center}
\includegraphics[width=8.5cm,clip]{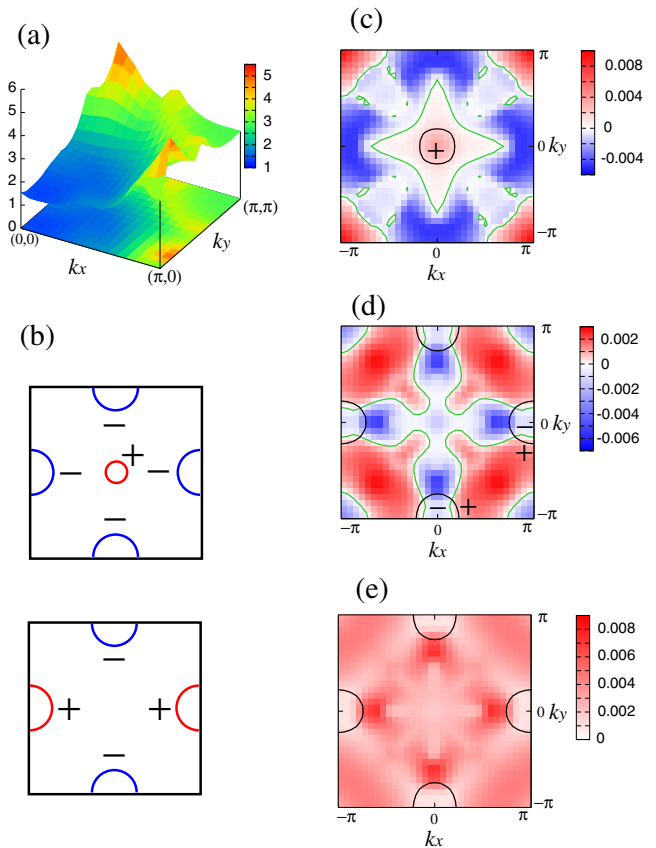}
\caption{(color online)
RPA result for the spin 
susceptibility $\chi_s$ (a), 
the gap functions $\phi_3$ (c) and  $\phi_4$ (d),  
$\sqrt{(\hat{\phi}\hat{\phi}^\dagger)_{44}}$ (e) 
for $U=1.2$, $U'=0.9$, $J=J'=0.15$, $n=6.1$ and $T=0.02$ (in eV).
In (c) and (d), 
the black (green) solid lines represent the Fermi surfaces (gap nodes).
In (b), the fully gapped extended $s$- (upper panel) and 
$d_{x^2-y^2}$-wave gaps are schematically shown.
\label{fig2}}
\end{center}
\end{figure}

Let us first look at the result for 
$\chi_s$ for $U=1.2$, $U'=0.9$, $J=J'=0.15$, and $T=0.02$ 
(all in units of eV) in Fig.\ref{fig2}(a). 
The spin susceptibility has peaks around 
$(k_x, k_y) = (\pi,0)$, $(0,\pi)$ and also a ridge-like structure 
from $(\pi,\pi/2)$ to $(\pi/2,\pi)$.
This in fact reflects the Fermi surface in Fig.\ref{fig1}(b), 
where we have two 
kinds of nesting vector:
$\sim (\pi,0),(0,\pi)$ across $\alpha$ and $\beta$, and 
 $\sim (\pi,\pi/2), (\pi/2,\pi)$
across $\beta_1$ and $\beta_2$.  A good nesting enhances tendency towards 
magnetism. In the RPA (where the self-energy correction
in the Green's function is neglected), 
we have to take $U$ as small as 1.2eV to ensure 
magnetic ordering does not take place in the temperature range 
considered.

For SC, we show in Fig.\ref{fig2}(c)(d)
the gap function for bands 3 and 4 (as counted from below), together
with the Fermi surface of each band. At this temperature ($T=0.02$), 
the eigenvalue of the Eliashberg equation is $\lambda=0.96$.\cite{comment3} 
The gap is basically $s$-wave, but 
changes sign between the Fermi surface  
of band 3 $(\alpha_2)$ (and also band 2; $\alpha_1$, not shown) 
and those of band 4 $(\beta_1,\beta_2)$, 
namely, across the nesting vector $\sim (\pi,0),(0,\pi)$ (M point in the 
original BZ) at which the spin fluctuations develop.
Such a sign change of the gap between inner hole and outer electron 
Fermi pockets 
is analogous to those in models studied by Bulut {\it et al.},\cite{Bulut} 
and also by two of the present authors.\cite{KA,KA2} 
It is also reminiscent of the unconventional $s$-wave pairing 
mechanism for Na$_x$CoO$_2\cdot y$H$_2$O\cite{Takada} 
proposed by four of the present authors.\cite{KKNCOO} 
After completion of the present study, we have come to notice that a 
recent preprint by Mazin {\it et al.} also  
concludes an $s$-wave pairing in which the 
gap changes sign between $\alpha$ and $\beta$ Fermi surfaces,\cite{Mazin}
as schematically shown in the upper panel of Fig.\ref{fig2}(b).
For the present set of parameter values, in addition to this sign change, 
we find that the nodes of the gap intersect the $\beta$ Fermi surface.
This is because the spin fluctuations due to the $\beta_1$-$\beta_2$ nesting 
favor a sign change in the gap between $\beta_1$ and $\beta_2$ Fermi 
surfaces. In fact, we have found that this nodal line moves out of the 
$\beta$ Fermi surface for the parameter values for which 
the spin fluctuations due to the $\beta_1$-$\beta_2$ nesting 
become less effective, e.g., for $U=U'$. 
In that case, the gap becomes closer to the upper panel of 
Fig.\ref{fig2}(b).\cite{Onarihoneycomb}

We have so far focused on the diagonal 
elements of the gap matrix in the band representation, but 
to be more precise, we have to consider the off-diagonal 
(interband) elements 
in order to make accurate comparison with experiments.
The off-diagonal elements in the present case turn out to be not 
negligibly small due to the heavy entanglement of the bands. 
One way to look at this effect is to calculate 
the quantity $\sqrt{(\hat{\phi}\hat{\phi}^\dagger)_{44}}$, 
where $\hat{\phi}$  is the 
gap matrix and 44 denotes the diagonal element of band 4. 
As shown in Fig.\ref{fig2}(e),  
we find that this quantity 
is finite over the entire BZ, but a remnant of the nodal lines of the 
diagonal element still 
appears as a valley that intersects the $\beta$ Fermi surface. 
In this sense, we can say that the magnitude of the gap varies 
along the $\beta$ Fermi surface 
(becomes large at points far from the BZ edge) 
if the spin fluctuations arising from 
$\alpha$-$\beta$ and $\beta_1$-$\beta_2$ interactions have competing 
strength. The degree of the variation of the gap in the actual materials 
may be determined experimentally from 
the density of states, e.g., tunneling spectroscopy, or 
directly by angle resolved photoemission studies.

In the above, we mainly considered the possibility of 
unconventional $s$-wave pairing. On the other hand, 
if the $\alpha$ Fermi surfaces are absent (or less effective), 
the simplest form of the gap 
would be the $d_{x^2-y^2}$-wave pairing ($d_{XY}$ in the original BZ), 
where the gap changes sign between $\beta_1$ and $\beta_2$ Fermi
surfaces as shown in the lower panel of Fig.\ref{fig2}(b). 
To check this, we have performed an RPA calculation on (i) the present model 
with $n=6.3$ and (ii) a model where 
we artificially shift the crystal 
field splitting to let the $\alpha$ Fermi surfaces
disappear for $n=6.1$. In both cases, we indeed obtain the $d_{x^2-y^2}$-wave.
Since the band structure generally changes from the LDA result due to 
correlation effects\cite{Kotliar} or 
a band filling different from the formally expected value, 
we leave, at the present stage, 
this $d$-wave state as another candidate for the pairing symmetry 
in this material.



Many other interesting problems remain for 
future studies. Spin fluctuations and SC should
be studied by taking into account the self-energy correction, for which 
a fluctuation exchange study is underway\cite{Onari} 
It is also intriguing to investigate whether the present 
unconventional gap can quantitatively account for 
the specific-heat\cite{Mu} and point-contact conductance\cite{Shan} results.
Also, further insight into the origin of the 
high $T_c$ SC in LaFeAsO$_{1-x}$F$_x$ may be 
obtained by performing similar microscopic studies on 
LaFePO$_{1-x}$F$_x$\cite{Hosono1} or LaNiPO.\cite{Hosono3}

We are grateful to Hideo Hosono for fruitful discussions and 
providing us with the lattice structure data prior to publication.
RA is grateful to Kazuma Nakamura for 
discussions on the Wannier orbitals.
Numerical calculations were performed at the facilities of
the Information Technology Center, University of Tokyo, 
and also at the Supercomputer Center,
ISSP, University of Tokyo. 
This study has been supported by 
Grants-in-Aid for Scientific Research from  MEXT of Japan and from 
the Japan Society for the Promotion of Science.
%



\begin{thebibliography}{99}
\bibitem[*]{arita}  Present Address: Department of Applied Physics, 
University of Tokyo, Hongo, Tokyo, 113-8656, Japan.
\bibitem{Hosono2}
Y. Kamihara {\it et al.}, J. Am. Chem. Soc. {\bf 130}, 3296 (2008).
\bibitem{Zimmer} 
B.I. Zimmer {\it et al.}, J. Alloys Compd. {\bf 229}, 238 (1995).
\bibitem{Quebe}
P. Quebe {\it et al.},
J. Alloys Compd. {\bf 302}, 70 (2000).
\bibitem{Hosono1}
Y. Kamihara {\it et al.}, J. Am. Chem. Soc. {\bf 128}, 10012 (2006).
\bibitem{Hosono3}
T. Watanabe {\it et al.}, Inorg. Chem. {\bf 46}, 7719 (2007).
\bibitem{Chen}
G.F. Chen {\it et al.}, unpublished (arXiv: 0803.0128)
\bibitem{Mu}
G. Mu {\it et al.}, Chin. Phys. Lett. {\bf 25}, 2221 (2008).
\bibitem{Shan} 
L. Shan {\it et al.}, unpublished (arXiv: 0803.2405)
\bibitem{Hu} C.-R. Hu, Phys. Rev. Lett. {\bf 72}, 1526 (1994).
\bibitem{Tanaka1} Y. Tanaka and S. Kashiwaya, Phys. Rev. Lett. {\bf 74}, 
3451 (1995).
\bibitem{Tanaka3} S. Kashiwaya and Y. Tanaka, Rep.Prog. Phys. {\bf 63}, 1641
(2000).
\bibitem{Lofwander} T. L\"{o}fwander {\it et al.}, 
Supercond. Sci. Technol. {\bf 14}, R53 (2001).
\bibitem{Zhu} X. Zhu {\it et al.}, unpublished (arXiv:0803.1288)
\bibitem{Lebegue} S. Leb\'{e}gue, Phys. Rev. B {\bf 75}, 035110 (2007).
\bibitem{Singh}
D.J. Singh and M.-H. Du, unpublished (arXiv: 0803.0429).
\bibitem{Kotliar}
K. Haule {\it et al.}, 
Phys. Rev. Lett. {\bf 100}, 226402 (2008).
\bibitem{Xu} 
G. Xu et al, 
Europhys. Lett. {\bf 82}, 67002 (2008).
\bibitem{Boeri}
L. Boeri {\it et al.}, 
unpublished (arXiv: 0803.2703).
\bibitem{pwscf} 
S. Baroni {\it et al.}, 
http://www.pwscf.org/.
Here we adopt the exchange correlation functional introduced by
J. P. Perdew {\it et al.}
[Phys. Rev. B {\bf 54}, 16533 (1996)], and the wave functions are expanded by 
plane waves up to a cutoff energy of 40 Ry.
10$^3$ $k$-point meshes are used
with the special points technique by 
H.J. Monkhorst and J.D. Pack [Phys. Rev. B {\bf 13}, 5188 (1976)].
\bibitem{MaxLoc} N. Marzari and D. Vanderbilt, Phys. Rev. B 
{\bf 56}, 12847 (1997); 
I. Souza {\it et al.}, 
Phys. Rev. B {\bf 65}, 035109 (2002).
The Wannier functions are generated by the code developed by
A. A. Mostofi {\it et al.}
(http://www.wannier.org/) 
for the energy window $-2$ eV $<\epsilon_k-E_F<$ 2.6eV,
where $\epsilon_k$ is the eigenenergy of the Bloch states
and $E_F$ the Fermi energy.
\bibitem{comment}  While an ambiguity exists in unfolding the BZ 
[if the sign of the hoppings $t(\Delta x,\Delta y;\mu, \nu)$ with 
$\Delta x+\Delta y$=odd (i.e., hoppings between A and B sites in a bipartite 
lattice) is changed, a band structure is reflected with respect to  
$|k_x|+|k_y|=\pi$], 
this is just a unitary transformation, and either way the five-band structure 
gives the same ten bands in the folded BZ as well as the same 
RPA results.
\bibitem{comment4} Quasi 2D systems are generally more favorable than 
3D for spin fluctuation mediated SC as shown by 
P. Monthoux and G.G. Lonzarich, Phys. Rev. B {\bf 59}, 14598 (1999),  
and R. Arita {\it et al.}, {\it ibid} {\bf 60}, 14585 (1999).
\bibitem{Yada} K. Yada and H. Kontani, J. Phys. Soc. Jpn. {\bf 74}, 2161 
(2005).
\bibitem{Takimoto} T. Takimoto {\it et al.}, 
Phys. Rev. B {\bf 69}, 104504 (2004).
\bibitem{comment3} The $\lambda$ value should be reduced if 
the self-energy corrections are considered as in FLEX.
\bibitem{Bulut} N. Bulut {\it et al.}, 
Phys. Rev. B {\bf 45}, 5577 (1992).
\bibitem{KA} K. Kuroki and R. Arita, Phys. Rev. B {\bf 64}, 024501 (2001).
\bibitem{KA2} K. Kuroki {\it et al.}, 
Phys. Rev. B {\bf 66}, 184508 (2002).
\bibitem{Takada} K. Takada, Nature {\bf 422}, 53 (2003).
\bibitem{KKNCOO} K. Kuroki {\it et al.}, Phys. Rev. B {\bf 73}, 184503 (2006). 
\bibitem{Mazin} I.I. Mazin  {\it et al.}, unpublished (arXiv:0803.2740).
\bibitem{Onari} S. Onari {\it et al.}, unpublished.
\bibitem{Bickers} N.E. Bickers {\it et al.}, 
Phys. Rev. Lett. {\bf 62}, 961 (1989).
\bibitem{Onarihoneycomb} The situation is reminiscent of a 3D 
case of disconnected Fermi surface studied in 
S. Onari {\it et al.}, Phys. Rev. B {\bf 65}, 184525 (2002), 
where the presence of intra- 
and inter-pocket nestings dominates the pairing symmetry.
\end{thebibliography}
\end{document}